\documentclass[preprint]{revtex4}
\usepackage{bm,color}
\usepackage{graphicx}
\usepackage{amsmath}

\begin{document}
\title{Thermally Driven Ratchet Motion of Skyrmion Microcrystal and Topological Magnon Hall Effect}

\author{M. Mochizuki$^{1,2}$}
\email{mochizuki@phys.aoyama.ac.jp}

\author{X. Z. Yu$^3$}

\author{S. Seki$^{2,3,4}$}

\author{N. Kanazawa$^5$}

\author{W. Koshibae$^3$}

\author{J. Zang$^6$}

\author{M. Mostovoy$^7$}

\author{Y. Tokura$^{3,4,5}$}

\author{N. Nagaosa$^{3,4,5}$}

\affiliation{$^1$ Department of Physics and Mathematics, Aoyama Gakuin University, Sagamihara, Kanagawa 229-8558, Japan}

\affiliation{$^2$ PRESTO, Japan Science and Technology Agency, Kawaguchi, Saitama 332-0012, Japan}

\affiliation{$^3$ RIKEN Center for Emergent Matter Science (CEMS), Wako, Saitama 351-0198, Japan}

\affiliation{$^4$ Department of Applied Physics, Quantum-Phase Electronics Center, 
The University of Tokyo, Bunkyo-ku Tokyo 113-8656, Japan}

\affiliation{$^5$ Department of Applied Physics, The University of Tokyo, Bunkyo-ku, Tokyo 113-8656, Japan}

\affiliation{$^6$ Department of Physics and Astronomy, Johns Hopkins University, Baltimore, Maryland 21218, USA}

\affiliation{$^7$ Zernike Institute for Advanced Materials, University of Groningen, Nijenborgh 
4, 9747 AG, Groningen, The Netherlands}

\maketitle
{\bf Spontaneously emergent chirality is an issue of fundamental importance across the natural sciences~\cite{Gardner90}. It has been argued that a unidirectional (chiral) rotation of a mechanical ratchet is forbidden in thermal equilibrium, but becomes possible in systems out of equilibrium~\cite{Feynman63}. Here we report our finding that a topologically nontrivial spin texture known as a skyrmion - a particle-like object in which spins point in all directions to wrap a sphere~\cite{Skyrme62} - constitutes such a ratchet. By means of Lorentz transmission electron microscopy (Lorentz TEM) we show that micron-sized crystals of skyrmions in thin films of Cu$_2$OSeO$_3$ and MnSi display a unidirectional rotation motion. Our numerical simulations based on a stochastic Landau-Lifshitz-Gilbert (LLG) equation suggest that this rotation is driven solely by thermal fluctuations in the presence of a temperature gradient, whereas in thermal equilibrium it is forbidden by the Bohr-van Leeuwen theorem~\cite{Bohr11,Leeuwen21}. We show that the rotational flow of magnons driven by the effective magnetic field of skyrmions gives rise to the skyrmion rotation, therefore suggesting that magnons can be used to control the motion of these spin textures.}

The formation of triangular arrays of skyrmions in chiral-lattice magnets under an applied magnetic field was theoretically predicted~\cite{Bogdanov89,Bogdanov94,Rossler06} and experimentally observed in MnSi~\cite{Muhlbauer09,Tonomura}, Fe$_{1-x}$Co$_x$Si~\cite{Munzer10,Pfleiderer10,YuXZ10N}, FeGe~\cite{YuXZ10M}, and Cu$_2$OSeO$_3$~\cite{Seki12a,Adams12,Seki12b}. 
The magnetization in such crystals is antiparallel to the magnetic field $\bm B$ at the center of each skyrmion and is parallel to $\bm B$ at its periphery (see Figs.1a-c). Recently, real-space images of nanosized skyrmions have been successfully obtained by the Lorentz TEM. While scanning temperatures and magnetic fields in the experiments, we have encountered a peculiar dynamical phenomenon, that is, in a wide temperature interval excluding only the lowest temperatures, micro-scale regions of the skyrmion crystal (SkX) show, in addition to Brownian motion, a clearly discernible unidirectional rotation. 

\begin{figure}[hp]
\includegraphics[scale=1.0]{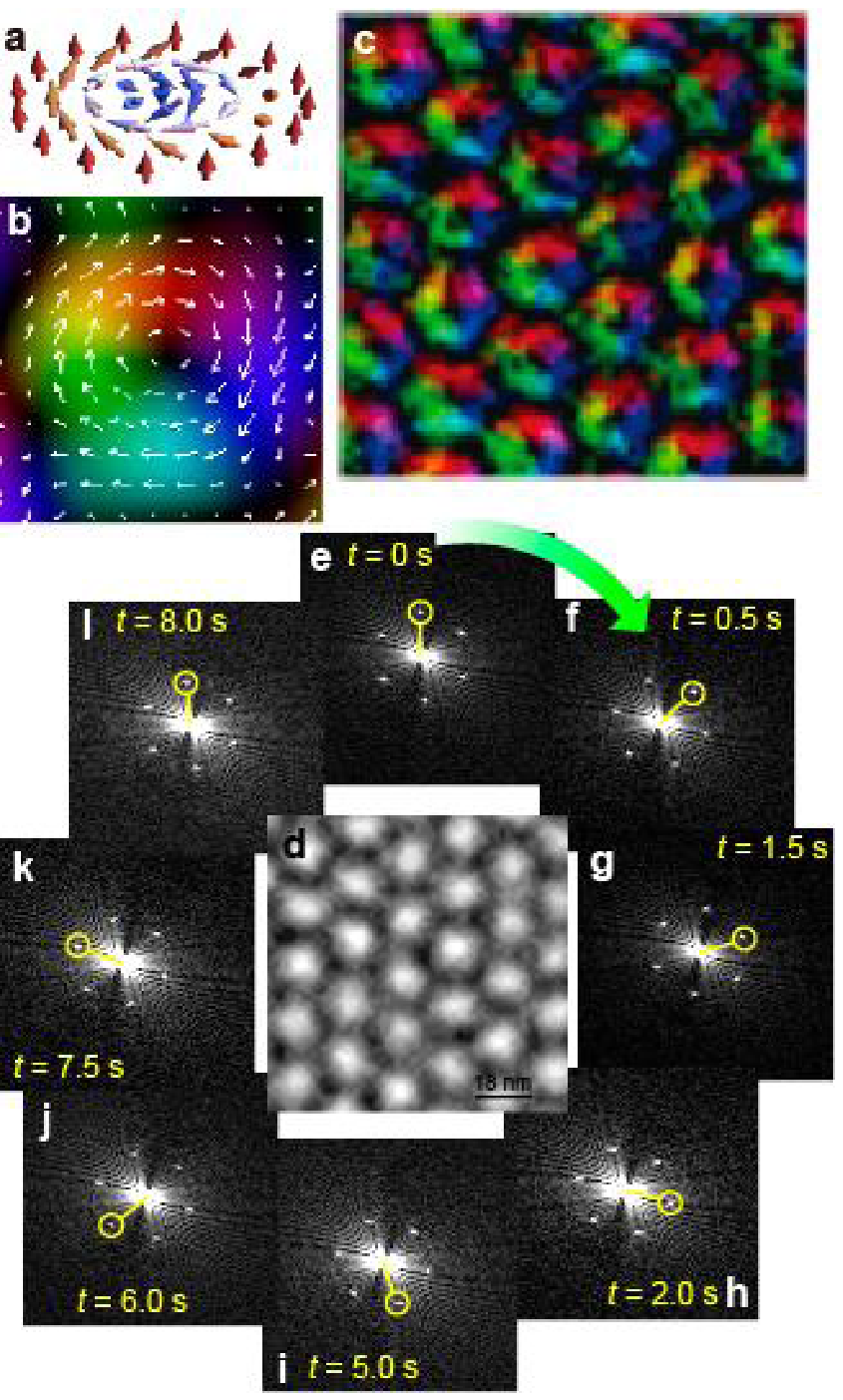}
\caption{}
\label{Fig1}
\end{figure}

In Supplementary Movies 1 and 2, we show examples of this phenomenon observed in thin-plate ($\sim$50nm-thick) specimens of MnSi and Cu$_2$OSeO$_3$, respectively. These compounds are chiral-lattice magnets with a common space group P2$_1$3. In zero field they undergo a phase transition from paramagnetic to helimagnetic phase at 29.5 K and 60 K, respectively. The helix period $\lambda_m$ is, respectively, 18 nm and 50 nm. Thin-plate ($<$100 nm-thick) specimens of these compounds host stable SkX phases with the skyrmion lattice spacing $\sim \frac{2}{\sqrt{3}}\lambda_m$. The field strength for the Movies 1 and 2 is 175 mT and 65 mT, respectively, and $\bm B$ is applied in the negative $z$-direction perpendicular to the plate plane: $\bm B \parallel -\bm z$. The observed (static) Lorentz-TEM image of SkX in MnSi specimen is displayed in Figs.~\ref{Fig1}c and d. In Figs.~\ref{Fig1}e-l, we exemplify time evolution (snapshots) of Fourier components of the magnetic configurations where a hexagon composed of six Bragg peaks rotates clockwise, indicating the clockwise rotation of the SkX  domains. 

The rotation rate depends on the irradiation density of the electron beam of the microscope (Supplementary Information). The skyrmion rotation is observed only above a critical irradiation density and the rotation rate grows as the density increases. This indicates that this rotation is a non-equilibirum phenomenon induced by the electron beam. The similarity between the chiral rotations of SkX micro-domains and their hexagon-shaped Fourier components observed in MnSi (Supplementary Movie 1) and Cu$_2$OSeO$_3$ (Supplementary Movie 2) is remarkable in view of different origins of magnetism in these two compounds (MnSi is a metal, while Cu$_2$OSeO$_3$ is an insulator) and the difference in skyrmion parameters, e.g. the transition temperature and the SkX lattice spacing. Hence, such a ratchet motion should be viewed as a generic feature of skyrmion systems. Then the question arises why and how it occurs.

A circular magnetic field induce by the electron beam of the Lorentz TEM is estimated to be five orders of magnitude smaller than the geomagnetic field, and thus cannot cause the rotation. Recently it was demonstrated that a spin-polarized electric current parallel to the sample plane can drive similar rotations of SkX domains in the presence of thermal gradient~\cite{Jonietz10,Everschor12}. In our case, however, a possible effect of electric currents can be excluded since the electron beam of Lorentz TEM is three orders of magnitude smaller than the threshold current of 10$^5$-$10^6$ A/m$^2$ for the current-driven skyrmion motions~\cite{Jonietz10,Everschor12,YuXZ12,Iwasaki13} and, in addition, its direction is perpendicular rather than parallel to the film, which further reduces spin-torque effects. Furthermore, the energy of the electrons is very high and the interaction with the spins in the specimen is very weak except through the magnetic field induced by the magnetization. This conclusion is corroborated by the fact that the skyrmion rotation is also observed in the insulating Cu$_2$OSeO$_3$. Thus, the unidirectional rotation is apparently induced by thermal effects.

In order to clarify the nature of this phenomenon, we performed numerical simulations. The SkX phase in chiral-lattice magnets is described by a classical Heisenberg model on the two-dimensional square lattice~\cite{YiSD09}, which contains ferromagnetic-exchange and Dzyaloshinskii-Moriya (DM) interactions as well as the Zeeman coupling to $\bm B$=(0, 0, $B_z$) normal to the plane~\cite{Bak80}. 
The Hamiltonian is given by,
\begin{eqnarray}
\mathcal{H}&=&-J \sum_{<i,j>} \bm m_i \cdot \bm m_j 
-D \sum_{i} (\bm m_i \times \bm m_{i+\hat{x}} \cdot \hat{x}
+\bm m_i \times \bm m_{i+\hat{y}} \cdot \hat{y}) \nonumber \\
&-&\bm B \cdot \sum_i \bm m_i,
\label{eqn:model}
\end{eqnarray}
where the magnetization vector $\bm m_i$ is defined as $\bm m_i=-\bm S_i/\hbar$ with the norm $m$=$|\bm m_i|$. The spin turn angle $\theta$ in the helical phase is determined by the ratio $D/J$ and for $D/J$=0.27 used here, $\theta$=11$^{\circ}$ corresponding to the period of $\sim$33 sites. With a typical lattice constant of 5 \AA, this gives $\lambda_m \sim$17 nm, which is comparable to the helical period $\sim$18 nm in MnSi (our conclusions, however, are not affected by the choice of the value of $D/J$). Our Monte-Carlo analysis shows that the SkX phase emerges in the range of $1.688\times 10^{-2} < |B_z|/Jm < 5.67\times 10^{-2}$ between the helical and ferromagnetic phases.

\begin{figure}[hp]
\includegraphics[scale=1.0]{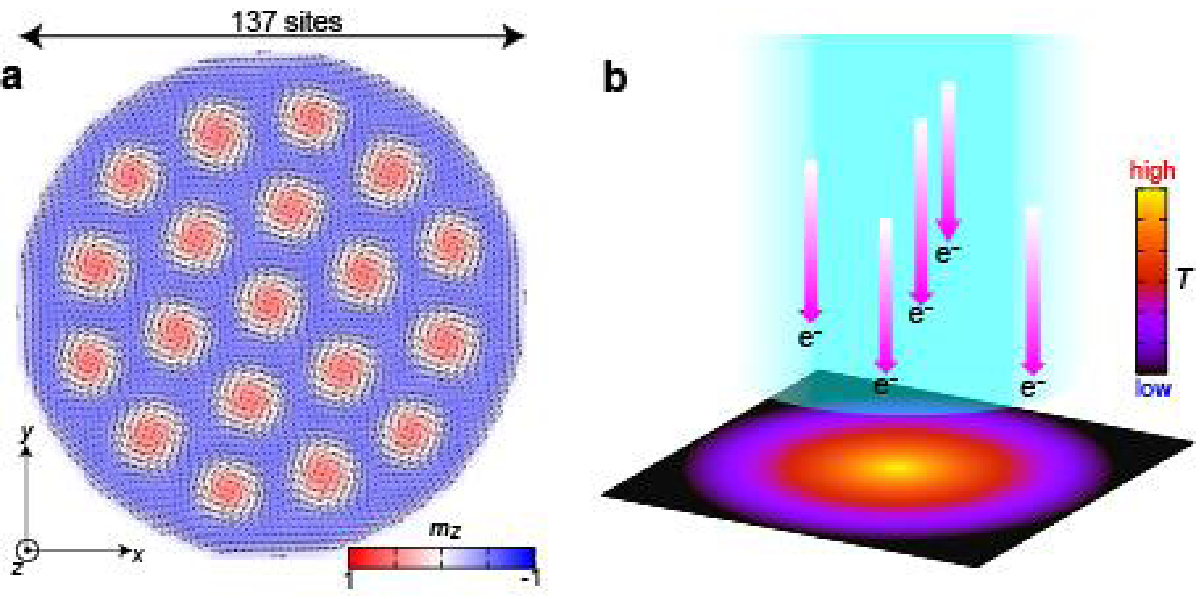}
\caption{}
\label{Fig2}
\end{figure}

In this study, we treat SkX confined in a micro-scale circular disk with the diameter $2R=137$ sites, as shown in Fig.~\ref{Fig2}a. We impose the open boundary condition and simulate thermally-induced dynamics of this skyrmion microcrystal by numerically solving a stochastic LLG equation~\cite{Brown63,Kubo70} using the Heun scheme~\cite{GPalacios98}. The equation is given by
\begin{eqnarray}
\frac{d \bm m_i}{d t}
=-\frac{1}{1+\alpha_{\rm G}^2} \left[
\bm m_i \times \left( \bm B^{\rm eff}_i + \bm \xi^{\rm fl}_i(t) \right)
+ \frac{\alpha_{\rm G}}{m} \bm m_i \times \left[
\bm m_i \times \left( \bm B^{\rm eff}_i + \bm \xi^{\rm fl}_i(t)
\right) \right] \right],
\label{eq:LLGEQ}
\end{eqnarray}
where $\alpha_{\rm G}$ is the Gilbert-damping coefficient and $\bm B^{\rm eff}_i = -\frac{1}{\hbar} \frac{\partial \mathcal{H}}{\partial \bm m_i}$ is the deterministic field. The Gaussian stochastic field $\bm \xi^{\rm fl}_i(t)$ describes effects of thermally fluctuating environment interacting with $\bm m_i$, which satisfies  $\left< \xi^{\rm fl}_{i,\beta}(t) \right>=0$ 
and $\left <\xi^{\rm fl}_{i,\beta}(t) \xi^{\rm fl}_{j,\lambda}(s) \right>=
2\kappa \delta_{ij}\delta_{\beta\lambda} \delta(t-s)$, where $\beta$ and $\lambda$ are Cartesian indices. The fluctuation-dissipation theorem gives a relation between $\kappa$ and temperature $T$: $\kappa=\alpha_{\rm G}k_{\rm B}T/m$~\cite{GPalacios98}. The initial spin configuration (Fig.~\ref{Fig2}a) is prepared by the Monte-Carlo thermalization at low temperature and by further relaxing it in the LLG simulation at $T$=0. Starting from this initial configuration, we generate the random numbers corresponding to the stochastic force $\bm \xi^{\rm fl}_i(t)$ and solve equation~(\ref{eq:LLGEQ}). In what follows we use units in which the lattice constant $a=1$, the exchange energy $J=1$, the Boltzmann constant $k_{\rm B}=1$ and $\hbar=1$.

In thermal equilibrium we only find Brownian motion of skyrmions and no unidirectional rotation. We then include a radial temperature gradient to examine whether it can give rise to chiral rotation. In the Lorentz-TEM experiment, the electron beam is irradiated onto a thin-plate specimen, which inevitably raises temperature of the beam spot with respect to the outer region, resulting in the temperature gradient as shown in Fig.~\ref{Fig2}{\bf b}. We consider a constant temperature gradient with $T$=$T_0$ at the edge to $T$=$T_0+\Delta T$ at the center in the circular-disk system, i.e., $-dT/dr$=$\Delta T/R$.

\begin{figure}[hp]
\includegraphics[scale=1.0]{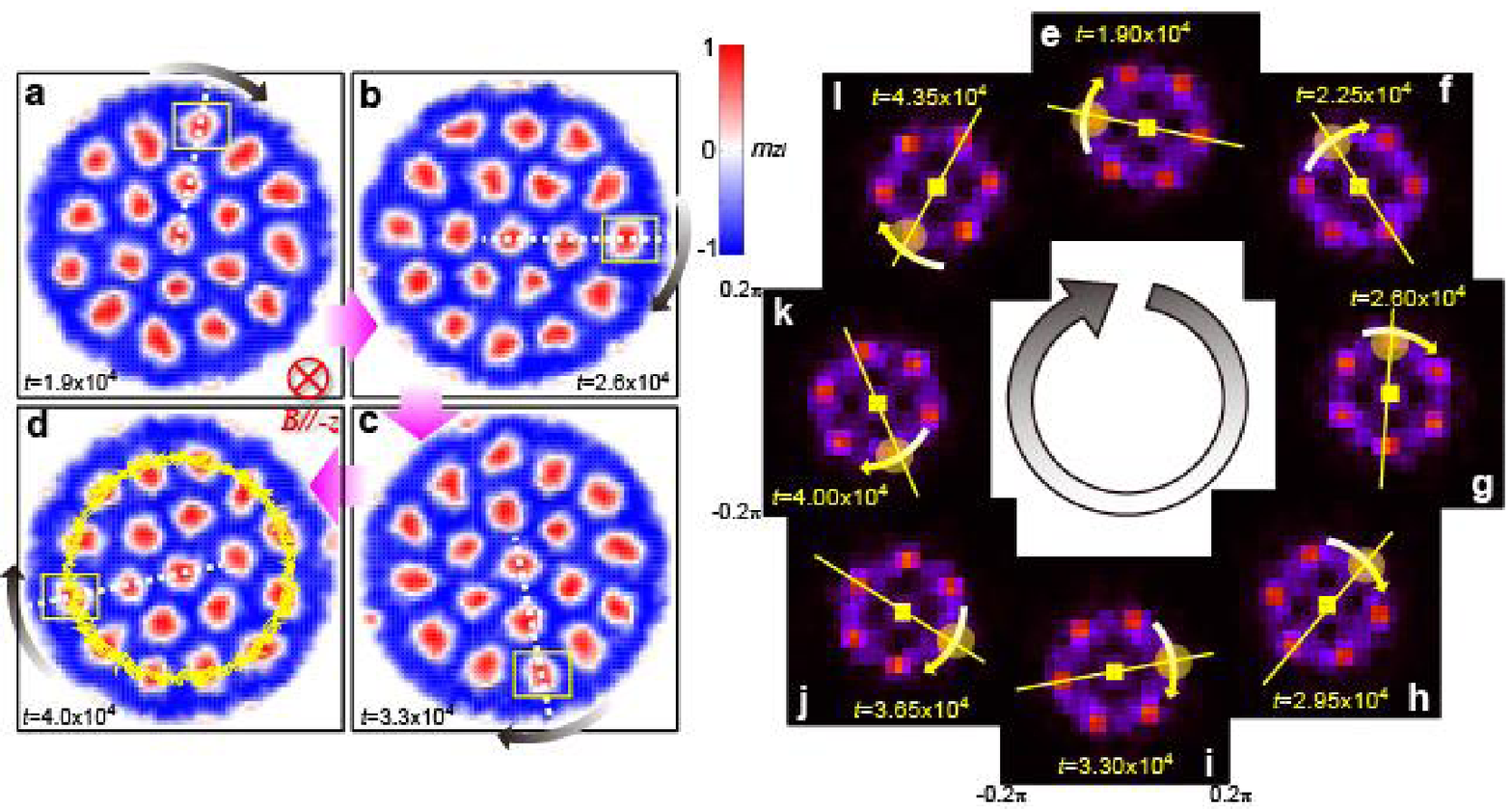}
\caption{}
\label{Fig3}
\end{figure}

We display snapshots of the calculated real-space magnetization dynamics at selected times in Figs.~\ref{Fig3}a-d, and the trajectory of a selected skyrmion indicated by solid rectangles in Fig.~\ref{Fig3}d. We find persistent rotation of SkX (Supplementary Movies 3 and 4). Figures~\ref{Fig3}e-l show the time evolution of Fourier transform of the spin structure - the rotating hexagon composed of six Bragg peaks (Supplementary Movie 5). In the simulation, we apply $\bm B \parallel -\bm z$ in accord with a setup of the Lorentz-TEM experiment, and find {\it clockwise} rotations in agreement with the experimental observations. Remarkably, this unidirectional rotation is driven purely by the thermal gradient because no other motive forces are considered in our simulation.

This nonreciprocal dynamics of SkX can be traced back to the algebra of spin operators, which determines the direction of spin precession in an applied $\bm B$. Equations of motion for the center-of-mass coordinates of a skyrmion have the form (Supplementary Information) 
\begin{eqnarray}
\left\{
\begin{array}{ccc}
\mathcal{M} \ddot{Y} + \alpha_{\rm G} \Gamma \dot{Y} + 4\pi QS \dot{X}
&=&-\frac{\partial U}{\partial Y} + 4\pi QJ^{\rm magnon}_x,\\
\mathcal{M} \ddot{X} + \alpha_{\rm G} \Gamma \dot{X} - 4\pi QS \dot{Y}
&=&-\frac{\partial U}{\partial X} - 4\pi QJ^{\rm magnon}_y,
\end{array}
\right.
\label{eqn:EqnMotion}
\end{eqnarray}
where $\mathcal{M}$ is the skyrmion mass, $Q$ is the topological charge ($Q=-1$ for $B_z<0$), $\alpha_{\rm G}$ is the Gilbert-damping constant, $\Gamma \approx 5.577\pi S$, $U$ is the external potential, and $\bm J^{\rm magnon}=(J^{\rm magnon}_x, J^{\rm magnon}_y)$ is the magnon current density defined in Supplementary Information, which drives skyrmion motion through the spin transfer torque. The extension to a many skyrmion problem is straightforward.

First we note that, if one replaces the magnon current by the stochastic Langevin force, the dynamics of skyrmions becomes equivalent to that of classical particles in an external magnetic field. In this case, the famous Bohr-van Leeuwen theorem~\cite{Bohr11,Leeuwen21}, which forbids orbital magnetism of classical particles in thermal equilibrium, precludes the spontaneous rotation of skyrmions, 
in agreement with our numerical results.

Next we discuss how a nonzero temperature gradient can induce a persistent rotation. One possible scenario is that the gradient $dT/dr$ modifies the radial distribution of skyrmions resulting in a non-vanishing radial force $-\left< dU/dr \right>$, which according to equation~(\ref{eqn:EqnMotion}) gives rise to a nonzero angular velocity. This, however, is forbidden, as the Bohr-van Leeuwen theorem can be generalized to the case of local equilibrium with a spatially inhomogeneous temperature $T(\bm r)$. What is required, is a nonequilibrium state with a heat flow forming a heat engine, in which an amount of heat $Q_1$ transferred from the high-temperature side ($T = T_1$) is partly transformed into work $W$, while the remaining heat $Q_2 = Q_1- W$ is absorbed on the low-temperature side ($T=T_2<T_1$). The ratchet rotation requires $W>0$ and the engine efficiency $\eta=W/Q_1$ is less than $1-T_2/T_1$.

The underlying microscopic mechanism involves the flow of the spin energy. There are two possible heat carriers: skyrmions and magnons. The dimensional analysis of the ratchet rotation frequency $\nu$ due to the thermal motion of skyrmions gives $\nu \sim \frac{1}{R} \left(-\frac{d T}{d r}\right)$ (in the $J=\hbar=k_{\rm B}=a=1$ units), which is too small to explain the results of numerical simulations. Magnons, on the other hand, are much more efficient in driving the rotation of skyrmions through the spin transfer torque~\cite{Kong13}. Importantly, a skyrmion induces an effective magnetic field $h_z=-{\bm m} \cdot (\partial_x {\bm m} \times \partial_y {\bm m})$ with the total flux $\int\! d{\bm r}h_z=4\pi Q$, which exerts the Lorentz force on magnons~\cite{Nagaosa12} (Supplementary Information). The skew scattering of magnons off skyrmions gives rise to the Topological Magnon Hall Effect: in addition to the thermally-driven magnon current in the radial direction, $J^{\rm magnon}_r =\kappa^{\rm magnon}_{xx} \left(-\frac{dT}{dr}\right)$, there is a current $J^{\rm magnon}_\theta = \kappa^{\rm magnon}_{xy} \left(-\frac{dT}{dr}\right)$ in the direction transverse to the temperature gradient corresponding to the {\it counterclockwise} rotation of the magnon gas.

\begin{figure}[hp]
\includegraphics[scale=1.0]{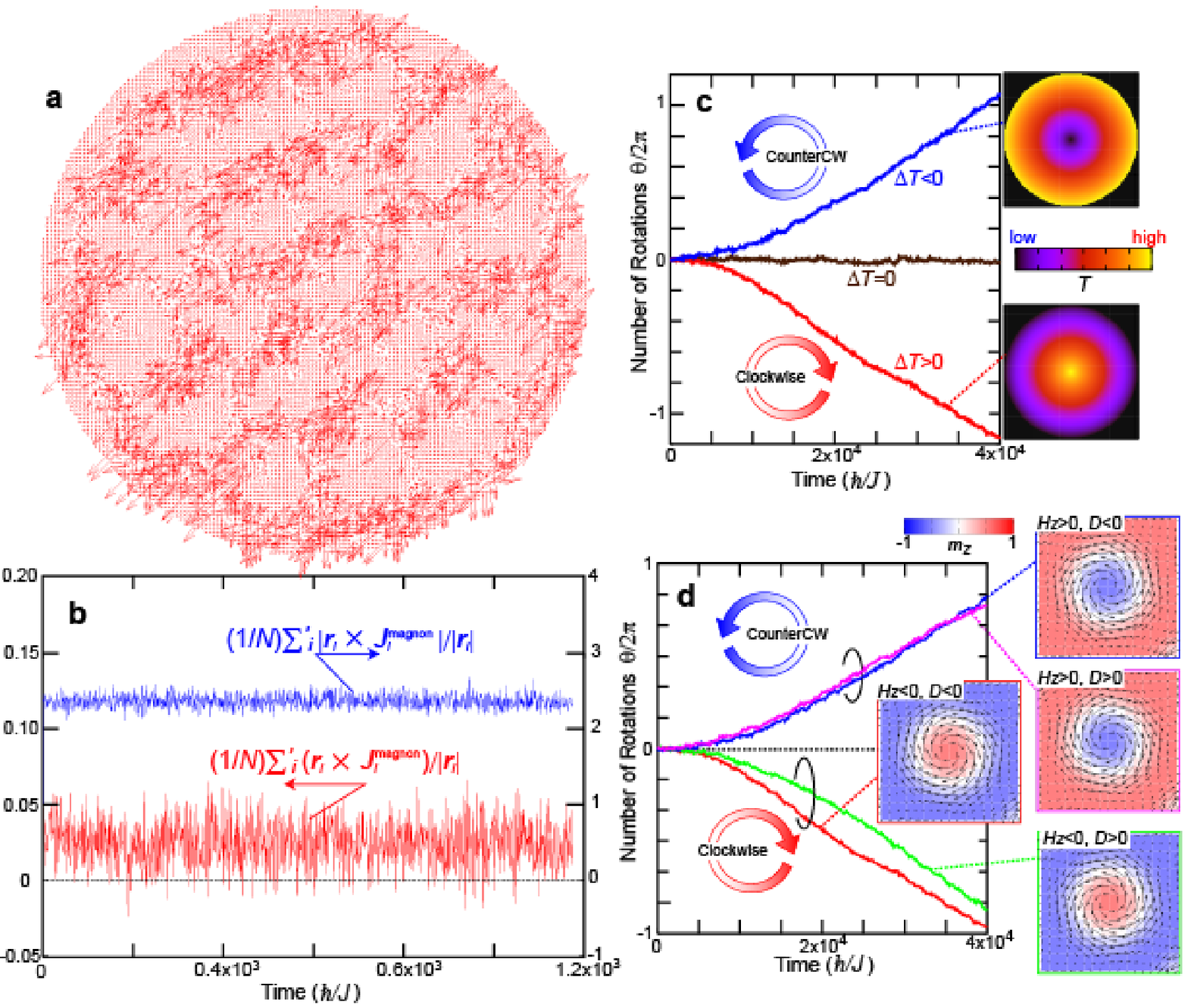}
\caption{}
\label{Fig4}
\end{figure}

Figure~\ref{Fig4} displays the result of simulations for the magnon current density. We show a real-space map of the magnon current density at a selected time in Fig.~\ref{Fig4}a where the arrows point in the current directions while their lengths represent the current amplitudes. In this map, the currents may seem to flow in random directions, but we find a net positive value of the time-averaged quantity $\left< J_\theta^{\rm magnon} \right> \equiv \frac{1}{N} \sum^{\prime}_i (\bm r_i \times \bm J_i^{\rm magnon})/|\bm r_i|$ shown in Fig.~\ref{Fig4}b. Here the vector $\bm r_i$ connects the center of the disk and the $i$th site, and the summation $\sum^{\prime}_i$ goes over the sites with $|\bm r_i|>20$. The positive sign indicates that the magnon current flows in the {\it counterclockwise} direction. The magnitude of the transverse magnon current in our numerical simulations, $J^{\rm magnon}_\theta \sim 10^{-2}$, which for $-dT/dr \sim 10^{-4}$ corresponds to a rather large magnon Hall conductivity, $\kappa^{\rm magnon}_{xy} \sim 10^{2}$~\cite{Hoogdalem13}.

The skew magnon scattering off skyrmions exerts a reaction force on skyrmions in the negative $\theta$-direction. The reaction force appears in the right-hand side of Eqs.(\ref{eqn:EqnMotion}), e.g., the force $F_y$ equals a product of the flux of the effective magnetic field $4\pi Q$ and the magnon current $J^{\rm magnon}_x$. From equations of motion (\ref{eqn:EqnMotion}), we obtain the estimate for the rotation rate of SkX (Supplementary Information)
\begin{eqnarray}
\nu \sim - \frac{J_\theta^{\rm magnon}}{\pi R} \sim - 5 \times 10^{-5},
\label{eqn:estimate}
\end{eqnarray}
for $J^{\rm magnon}_\theta \sim 10^{-2}$. The minus sign corresponds to {\em clockwise} rotation of skyrmions. Equation (\ref{eqn:estimate}) shows that the experimentally observed clockwise rotation of skyrmions is a consequence of the anticlockwise rotation of magnons. The estimate (\ref{eqn:estimate}) is in good quantitative agreement with the numerical result for the SkX rotation rate, $-3 \times 10^{-5}$.

We thus showed that the magnon current induced by the temperature gradient is deflected by the emergent magnetic field of skyrmions, which in turn gives rise to the rotation of SkX through the spin-transfer torque. Since the sign of $\bm J^{\rm magnon}$ is governed by the sign of $dT/dr$, the rotation of SkX should be reversed upon the sign reversal of the temperature gradient, which is indeed what we find in our simulation (see Fig.~\ref{Fig4}c and Supplementary Movie 6). Also the rotation direction becomes reversed upon the sign reversal of magnetic field $B_z$ but not upon the sign reversal of the DM parameter $D$ as seen in Fig.~\ref{Fig4}d because the former changes the sign of $\bm J^{\rm magnon}$ but the latter does not. This shows that skyrmion-magnon interactions and thermal spin fluctuations provide a key to understanding of the observed chiral rotation of skyrmions.

The proposed physics behind the observed rotation is distinct from the Skyrmion Hall effect discussed in a recent paper by Kong and Zang~\cite{Kong13}. They theoretically proposed that a longitudinal skyrmion motion due to the magnon current along the thermal gradient is accompanied by a small transverse motion due to the Gilbert damping. This Skyrmion Hall effect necessarily requires the longitudinal motion. In our case, however, the skyrmion motion in the radial direction (parallel to the temperature gradient) is forbidden due to the geometrical confinement. Hence, no Skyrmion Hall effect is possible and the topological magnon Hall effect is the only source of the observed skyrmion rotation. The reaction force from the magnon current deflected by the effective magnetic field of the topological skyrmion texture drives the peculiar chiral motion.

Note that the time scale of the rotation is microseconds in the simulation for $J$$\sim$1 meV, while it is a few seconds in the experiment. This discrepancy can be related to the strong sensitivity of the rotation rate to the shape of the boundary of the SkX, the magnitude of the temperature gradient and sample inhomogeneities, such as impurities and defects. Thus we observe no rotation in a rectangle-shaped system within a realistic simulation time (limited by a few milliseconds), apparently because the large friction between the SkX and the system edges makes the rotation rate extremely slow. In the circular-disk system, the rotation rate decreases as $\Delta T$ decreases, and vanishes when $\Delta T$=0. In addition, the rotation is less pronounced and its rate is lower in a larger-sized disk, indicating the absence of rotation in the thermodynamic limit.

To summarize, we have found experimentally and explained theoretically that micron-sized skyrmion crystals behave as the Feynman's ratchet. In the presence of the radial magnon flow, skyrmions exhibit persistent rotation in the direction determined by the sign of the topological charge of the skyrmions. The physical origin of this unusual phenomenon can be traced back to the chiral nature of spin dynamics. Our finding shows how thermal spin fluctuations can be harnessed to control topological spin textures by irradiating them with light and electrons. The manipulation of skyrmions with magnons can be used to build all-spin memory and logic devices with low dissipation losses by replacing the charge current by magnon current especially in the insulating magnets~\cite{Kruglyak10}.

\section*{Methods Summary}
The single crystal samples of MnSi were grown by the floating zone technique, while those of Cu$_2$OSeO$_3$ were grown by the chemical vapor transport method. For the real-space imaging of spin textures, their thin specimens with thickness of $\sim$50 nm were prepared by mechanical polishing and subsequent argon-ion thinning with an acceleration voltage of 4 kV. All experiments were performed with a transmission electron microscope (JEM2100F, JEOL) at an acceleration voltage of 200 kV. Images of the SkX were obtained at the over-focused Lorentz-TEM mode. The movies of Lorentz-TEM image were taken with the exposure time of 50 ms and the frame rate of 18 fps (frame per second). A liquid Helium cooling holder was utilized to investigate the $T$ dependence, by which $T$ at the specimen can be controlled from 6 K to 300 K. The electron beam strength is $2.7\times 10^3$ Am$^{-2}$ ($3.6\times 10^2$ Am$^{-2}$) for the MnSi (Cu$_2$OSeO$_3$) specimen. The magnetic field of $B$=175 mT (65 mT) parallel to the electron beam was applied, and the measurement was performed at 8 K (40 K) for MnSi (Cu$_2$OSeO$_3$).


\section*{Acknowledgements}
\noindent
The authors thank A. Rosch, M. Ichikawa, Y. Matsui, Y. Ogimoto, and E. Saito for discussions. XZY is grateful K. Nishizawa and T. Kikitsu for providing a transmission electron microscope (JEM2100F). This research was in part supported by JSPS KAKENHI (Grant Numbers 24224009, 25870169, and 25287088), by the Funding Program for World-Leading Innovative R\&D on Science and Technology (FIRST Program), Japan, and by G-COE Program ``Physical Sciences Frontier" from MEXT Japan. M. Mostovoy was supported by FOM grant 11PR2928 and the Niels Bohr International Academy. JZ is supported by the Theoretical Interdisciplinary Physics and Astrophysics Center and by the U.S. Department of Energy, Office of Basic Energy Sciences, Division of Materials Sciences and Engineering under Award DEFG02-08ER46544.

\section*{Author Contributions}
\noindent
M. Mochizuki carried out the numerical simulations and analyzed the simulation data. XY carried out the Lorentz transmission electron microscopy measurement and analyzed the experimental data. SS carried out the crystal growth of Cu$_2$OSeO$_3$. NK carried out the crystal growth of MnSi. The whole work has been lead by NN and YT. The results were discussed and interpreted by M. Mochizuki, XY, WK, JZ, M. Mostovoy, YT, and NN. M. Mochizuki, M. Mostovoy, YT, and NN wrote the draft.

\section*{Competing Financial Interests}
\noindent
The authors declare that they have no competing financial interests.

\newpage
\section*{Figure Legends}
\noindent
{\bf Fig.1 $|$ Experimentally observed Skyrmions and unidirectional rotation of micro-scale skyrmion-crystal domains in MnSi.} {\bf a,} Schematic figure of skyrmion in chiral-lattice magnets. {\bf b,} Real-space Lorentz-TEM image of single skyrmion in MnSi. The colour map and arrows represent in-plane components of magnetizations. {\bf c,} Real-space Lorentz-TEM image of the static skyrmion crystal in MnSi in which skyrmions are packed to form a triangular lattice. {\bf d,} Over-focused Lorentz-TEM image of the skyrmion crystal in MnSi. {\bf e-l,} Time profiles of Fourier transforms of temporally changing magnetic structures in the skyrmion crystal observed in MnSi by the Lorentz TEM, which show a {\it clockwise} rotation. Open circles and solid lines are guides for the eyes.\\

\noindent
{\bf Fig.2 $|$ Setup of the numerical simulation.} {\bf a,} Magnetic configuration of the skyrmion microcrystal confined in a circular-shaped disk at $T$=0 where the in-plane magnetization components at sites ($i_x$, $i_y$) are indicated by arrows when mod($i_x$, 2)=mod($i_y$, 2)=0, while distribution of the magnetization $z$-axis components, $m_{zi}$, is shown by a color map. {\bf b,} Schematics of induced thermal gradient by the electron-beam irradiation in the Lorentz TEM experiment.\\

\noindent
{\bf Fig.3 $|$ Simulated thermally driven rotation of the skyrmion microcrystal.} {\bf a-d,} Snapshots of simulated temporally changing distribution of the magnetization $z$-axis components $m_{zi}$ at $t$=1.9$\times$10$^4$ ({\bf a}), $t$=2.6$\times$10$^4$ ({\bf b}), $t$=3.3$\times$10$^4$ ({\bf c}), and $t$=4.0$\times$10$^4$ ({\bf d}), which show a {\it clockwise} rotation. Here the time unit is $\hbar/J$. Also shown in {\bf d} is the trajectory of a selected skyrmion indicated by rectangles. {\bf e-l,} Time evolution of Fourier transforms of the magnetic structure in the reciprocal space $-0.2\pi<k_x<0.2\pi$ and $-0.2\pi<k_y<0.2\pi$ for $m$=1, $B_z/J=-0.03$, $\alpha_{\rm G}$=0.01, $k_{\rm B}T_0/J$=0.1, and $k_{\rm B}\Delta T/J$=0.006 at selected times between $t$=1.90$\times$10$^4$ and $t$=4.35$\times$10$^4$ with constant time intervals of $\Delta t$=0.35$\times$10$^4$, which also show {\it clockwise} rotation.\\

\noindent
{\bf Fig.4 $|$ Simulation of the magnon current density.} {\bf a,} Spatial distribution of the magnon current density $\bm J_i^{\rm magnon}$ at a selected time where lengths of the arrows are proportional to the local amplitudes of $\bm J^{\rm magnon}$. {\bf b,} Time evolution of the quantity $\frac{1}{N} \sum^{\prime}_i (\bm r_i \times \bm J_i^{\rm magnon})/|\bm r_i|$, which shows a finite and positive component of $\bm J_i^{\rm magnon}$ in the circumferential direction. Here $\bm r_i$ is the vector connecting the disk center and the $i$th site, and the summation $\sum^{\prime}_i$ runs over the sites with $|\bm r_i|>20$. The sum of the absolute values, that is, $\frac{1}{N} \sum^{\prime}_i |\bm r_i \times \bm J_i^{\rm magnon}|/|\bm r_i|$ is plotted to show that the deviation of the above quantity from zero is meaningful. The calculation is performed for $m$=1, $B_z/J=-0.03$, $\alpha_{\rm G}$=0.01, $k_{\rm B}T_0/J$=0.1, and $k_{\rm B}\Delta T/J$=0.006. {\bf c,} Simulated number of rotations $\theta/2\pi$ as functions of time in the cases of thermal equilibrium ($\Delta T=0$), positive $T$ gradient ($\Delta T>0$), and negative $T$ gradient ($\Delta T<0$). Here the negative (positive) slope of the plot means a clockwise (counterclockwise) rotation of the SkX. The result shows that the rotation sense is reversed upon the sign reversal of $\Delta T$. {\bf d,} Simulated $\theta/2\pi$ as functions of time for four kinds of combinations of signs of $B_z$ and the DM parameter $D$ for $k_{\rm B}\Delta T/J$=+0.006. The rotation sense changes depending on the sign of $B_z$ but not on the sign of $D$.

\newpage
\newpage
\newpage
\newpage


\begin{thebibliography}{999}
\bibitem{Gardner90}Gardner, M. The New Ambidextrous Universe (Freeman, 1990).

\bibitem{Feynman63} Feynman, R. P. The Feynman Lectures on Physics, Vol. 1. Chapter 46 (Addison-Wesley, 1963).

\bibitem{Skyrme62}Skyrme, T. H. R., A unified field theory of mesons and baryons. Nucl. Phys. {\bf 31}, 556-569 (1962).

\bibitem{Bohr11}Bohr, N. Studier over Metallernes Elektrontheori (Kobenhavns Universitet, 1911).

\bibitem{Leeuwen21}van Leeuwen, H. J. Problemes de la theorie electronique du magnetisme. Journal de Physique et le Radium 2 (12): 361-377 (1921).
\bibitem{Bogdanov89}Bogdanov, A. N. \& Yablonski\^i, D. A. Thermodynamically stable ``vortices" in magnetically ordered crystals: The mixed state of magnets. {\it Sov. Phys. JETP} {\bf 68}, 101-103 (1989).

\bibitem{Bogdanov94}Bogdanov, A. \& Hubert, A. Thermodynamically stable magnetic vortex states in magnetic crystals. {\it J. Mag. Mag. Mat.} {\bf 138}, 255-269 (1994).

\bibitem{Rossler06}R\"o{\ss}ler, U. K., Bogdanov, A. N. \& Pfleiderer, C. Spontaneous skyrmion ground states in magnetic metals. {\it Nature} {\bf 442}, 797-801 (2006).
\bibitem{Muhlbauer09}M\"uhlbauer, S. et al. Skyrmion lattice in a chiral magnet. {\it Science} {\bf 323}, 915-919 (2009).

\bibitem{Tonomura} Tonomura, A., et al. 
Real-Space Observation of Skyrmion Lattice in Helimagnet MnSi Thin Samples. {\it Nano Lett.} {\bf 12}, 1673-1677 (2012).

\bibitem{Pfleiderer10}Pfleiderer, C. et al. Skyrmion lattices in metallic and semiconducting B20 transition metal compounds. {\it J. Phys. Condens. Matter} {\bf 22}, 164207 (2010).

\bibitem{Munzer10}M\"unzer, A. et al. Skyrmion lattice in the doped semiconductor Fe$_{1-x}$Co$_x$Si. {\it Phys. Rev. B} {\bf 81}, 041203(R) (2010).

\bibitem{YuXZ10N}Yu, X. Z. et al. Real-space observation of a two-dimensional skyrmion crystal. {\it Nature} {\bf 465}, 901-904 (2010).

\bibitem{YuXZ10M}Yu, X. Z. et al. Near room-temperature formation of a skyrmion
crystal in thin-films of the helimagnet FeGe. {\it Nature Mater.} {\bf 10}, 106-109 (2010).

\bibitem{Seki12a}Seki, S., Yu, X. Z., Ishiwata, S., Tokura, Y. Observation of skyrmions in a multiferroic material. {\it Science} {\bf 336}, 198-201 (2012).

\bibitem{Adams12}Adams, T. et al. Long-wavelength helimagnetic order and skyrmion lattice phase in Cu$_2$OSeO$_3$. {\it Phys. Rev. Lett.} {\bf 108}, 237204 (2012).

\bibitem{Seki12b}Seki, S. et al. Formation and rotation of skyrmion crystal in the chiral-lattice insulator Cu$_2$OSeO$_3$. {\it Phys. Rev. B} {\bf 85}, 220406 (2012).


\bibitem{YiSD09}Yi, S. D., Onoda, S., Nagaosa, N. \& Han, J. H. Skyrmions and anomalous Hall effect in a Dzyaloshinskii-Moriya spiral magnet. {\it Phys. Rev. B} {\bf 80}, 054416 (2009).

\bibitem{Bak80}Bak, P. \& Jensen, M. H. Theory of helical magnetic structures and phase transitions in MnSi and FeGe. {\it J. Phys. C} {\bf 13}, L881-L885 (1980).

\bibitem{Jonietz10}Jonietz, F. et al. Spin transfer torques in MnSi at ultralow current densities. {\it Science} {\bf 330}, 1648-1651 (2010).

\bibitem{Everschor12}Everschor, K. et al. Rotating skyrmion lattices by spin torques and field or temperature gradients. {\it Phys. Rev. B} {\bf 86}, 054432 (2012).

\bibitem{YuXZ12}Yu, X. Z. et al. Skyrmion flow near room temperature in an ultralow current density. {\it Nature Commun.} {\bf 3}, 988 (2012).

\bibitem{Iwasaki13}Iwasaki, J., Mochizuki, M. \& Nagaosa, N. Universal current-velocity relation of skyrmion motion in chiral magnets. {\it Nature Commun.} {\bf 4,} 1463 (2013).
\bibitem{Brown63}Brown, Jr., W. F. Thermal fluctuations of a single-domain particle. {\it Phys. Rev.} {\bf 130}, 1677-1686 (1963).

\bibitem{Kubo70}Kubo, R. \& Hashitsume, N. Brownian motion of spins. {\it Prog. Theor. Phys. Suppl.} {\bf 46}, 210-220 (1970).

\bibitem{GPalacios98}Garc\'ia, J. L. \& L\'azaro, F. J. Langevin-dynamics study of the dynamical properties of small magnetic particles. {\it Phys. Rev. B} {\bf 58}, 14937-14958 (1998).

\bibitem{Kong13}Kong, L. \& Zang, J. Dynamics of an Insulating Skyrmion under a Temperature Gradient. {\it Phys. Rev. Lett.} {\bf 111}, 067203 (2013).

\bibitem{Nagaosa12}Nagaosa, N. \& Tokura, Y. Emergent electromagnetism in solids. {\it Phys. Scr. T}{\bf 146}, 014020 (2012). 

\bibitem{Hoogdalem13}van Hoogdalem, K. A., Tserkovnyak, Y. \& Loss, D. Magnetic texture-induced thermal Hall effects. {\it Phys. Rev. B} {\bf 87}, 024402 (2013).

\bibitem{Kruglyak10}Kruglyak, V. V., Demokritov, S. O. \& Grundler, D. Magnonics. {\it J. Phys. D: Appl. Phys.} {\bf 43}, 260301 (2010).
\end{thebibliography}
\end{document}